\documentclass[aps,prb,twocolumn,showpacs,10pt,floatfix,superscriptaddress]{revtex4-2}
\usepackage[colorlinks=true, linkcolor=blue, citecolor=blue, urlcolor=blue]{hyperref}
\usepackage{epsf}
\usepackage{bm}
\usepackage{hyperref}
\usepackage{amsfonts}
\usepackage{amssymb}
\usepackage{amsmath,mathtools}
\usepackage{array}
\usepackage{enumerate,dsfont}
\usepackage{dcolumn,multirow}
\usepackage{graphicx}
\usepackage[caption=false]{subfig}
\usepackage[utf8]{inputenc}
\usepackage{latexsym}
\usepackage{braket}
\usepackage{xcolor}
\usepackage{dcolumn}
\usepackage{siunitx}

\newcommand{\gat}{\Tilde{\gamma}}
\newcommand{\et}{\Tilde{\eta}}
\newcommand{\gt}{\Tilde{g}}

\newcommand{\lk}{\left(}
\newcommand{\rk}{\right)}

\newcommand{\be}{\begin{equation}}
\newcommand{\ee}{\end{equation}}

\begin{document}

\title{Ultrasound detection of emergent photons in generic quantum spin ice}

\author{Sophia Simon}
\affiliation{Department of Physics and Centre for Quantum Materials, University of Toronto, Toronto, Ontario, Canada M5S 1A7}
\author{Adarsh S. Patri}
\affiliation{Department of Physics and Centre for Quantum Materials, University of Toronto, Toronto, Ontario, Canada M5S 1A7}
\affiliation{Department of Physics, Massachusetts Institute of Technology, Cambridge, MA 02142, USA}
\author{Yong Baek Kim}
\affiliation{Department of Physics and Centre for Quantum Materials, University of Toronto, Toronto, Ontario, Canada M5S 1A7}
\date{\today}

\begin{abstract}

Experimental identification of quantum spin ice (QSI), a U(1) quantum spin liquid on the pyrochlore lattice hosting emergent photons, is a major challenge in frustrated magnets. In this work, we propose ultrasound measurements as a tool for probing the emergent photons of various QSI phases. Our analysis includes QSI phases in non-Kramers doublet compounds such as $\rm{Pr}_2 \rm{Zr}_2 \rm{O}_7$ as well as
dipolar-octupolar Kramers doublet compounds such as $\rm{Ce}_2 \rm{Zr}_2 \rm{O}_7$. The latter may host emergent photons associated with an octupolar component which renders them difficult to detect with inelastic neutron scattering.
We demonstrate theoretically how the speed of the emergent photons can be obtained from the renormalization of the phonon spectrum and show that ultrasound measurements provide a means of distinguishing the dipolar from the octupolar QSI phase in dipolar-octupolar materials. 

\end{abstract}
\maketitle

\section{Introduction}

While quantum spin ice (QSI), a quantum spin liquid (QSL) on the pyrochlore lattice with emergent U(1) gauge structure \cite{Gingras_QSI_review, Balents_Spin_liquids_2010, Savary_QSL_review, Banerjee_Hard-Core_Bosons_2008, Shannon_QSI_QMC_2012}, has been the subject of intense research efforts over the last decades, conclusive experimental evidence confirming the existence of QSI is still missing. In principle, it is sufficient to establish the absence of long-range magnetic order down to the lowest temperatures and additionally detect the exotic excitations associated with QSI, including gapless emergent photons as well as gapped magnetic monopoles and electric charges (spinons) \cite{Hermele_Photons_QSI_2004, Benton_Photons_QSI_2012, Chen_Spinons_QSI_2017, Mandal_E-Field_QSI}. 
However, unambiguous identification of such exotic excitations has been a major challenge in condensed matter experiments \cite{Knolle_Field_guide_SL, Wen_Exp_identification_QSL}.

In this work, we consider the issue of detecting the emergent photons in QSI. Inelastic neutron scattering, the standard method for probing excitations in spin systems, suffers from a vanishing intensity as the photon energy approaches zero \cite{Savary_gMFT}, which renders it a highly challenging task to identify the existence of the emergent photons.
Other proposals, such as thermal conductivity measurements \cite{Tokiwa_Thermal_conductivity_Photons}, may indicate the existence of emergent photons but require further investigations since, for example, phonon contributions associated with the intrinsic disorder present in these materials may lead to similar experimental signatures \cite{Rau_Frustrated_pyrochlore_2018}.
Here, we propose ultrasound measurements as a tool for characterizing the emergent photons in QSI. 
More specifically, we derive the renormalization of the phonon spectrum due to the emergent photons and show how to extract the speed of the photons from the renormalized speed of sound. 

The basic QSI Hamiltonian takes the form of a frustrated Ising model on the pyrochlore lattice (see Fig.~\ref{fig_pyrochlore}) with additional perturbative transverse terms generating quantum fluctuations:
\begin{equation}
    \mathcal{H}_{QSI} = \sum_{\langle i,j \rangle} J_{\parallel} S_{\parallel}^i S_{\parallel}^j + \text{transverse terms}.
\label{QSI_ham}
\end{equation}
The sum runs over nearest neighbors on the pyrochlore lattice, $J_{\parallel} > 0$ is the Ising coupling constant and $S_{\parallel}^i$ denotes the Ising component of a pseudospin-1/2 operator at site $i$.
Using the parton construction developed in \cite{Savary_gMFT, Lee_Generic_QSI}, one can establish a mapping between the pseudospin-1/2 operators and lattice electrodynamics by relating the Ising pseudospin component to an emergent electric field $\mathbf{E}$ and the transversal components to spinon bilinears dressed with the emergent photon. In the continuum limit, the low-energy theory of the Hamiltonian in Eq.~(\ref{QSI_ham}) can then be expressed in terms of the imaginary time quantum electrodynamics (QED) action
\begin{equation}
\mathcal{S}_{QED} = \int_{\tau, \textbf{r}} \lk \frac{1}{2K} (\partial_{\tau} \textbf{A})^2 +  \frac{U}{2} (\nabla \times \textbf{A})^2\rk.
\label{S_QED}
\end{equation}
Here, $\textbf{A}$ denotes the vector potential and we choose the gauge where the scalar potential $\phi = 0$ \cite{Hermele_Photons_QSI_2004, Savary_gMFT}. $K$ and $U$ are phenomenological constants determining the speed $v = \sqrt{UK}$ of the emergent photon.

The coupling of the pseudospins (or the emergent gauge fields in Eq.~(\ref{S_QED})) with the lattice degrees of freedom encourages ultrasound measurements as a keen probe to identify the existence of the emergent photon.
For sufficiently low energies, only the coupling of the Ising component to lattice degrees of freedom needs to be taken into account since the transversal pseudospin components involve gapped spinons.
Indeed, the form of the coupling is constrained by the symmetries of the pyrochlore lattice.
For instance, in the conventional setting, where the pseudospins arise from effective spin-1/2 Kramers doublets, all pseudospin components transform like dipole components. This leads to the standard ``dipolar'' QSI \cite{Rau_Frustrated_pyrochlore_2018, Gingras_QSI_review}, which may be realized in $\rm{Yb}_2 \rm{Ti}_2 \rm{O}_7$ \cite{Hodges_Yb2Ti2O7_2001, Ross_QSI_Exp_2011, Chang_Yb2Ti2O7_2012, Pan_Yb2Ti2O7_2014, Pan_Yb2Ti2O7_2015, Gaudet_Yb2Ti2O7_2016, Yaouanc_Yb2Ti2O7_2016, Thompson_Yb2Ti2O7_2017, Scheie_Yb2Ti2O7_2017, Scheie_Yb2Ti2O7_2020}.

Strong spin-orbit coupling and crystalline electric fields, however, can equip the pseudospins with more unusual transformation properties \cite{Kusunose_Multipoles, Kuramoto_Multipoles, Witczak_SOC_2014, Rau_SOC_2016, Schaffer_SOC_2016}. Inelastic neutron scattering experiments on $\rm{Ce}_2 \lk \rm{Sn}, \rm{Zr} \rk_2 \rm{O}_7$ suggest the crystal field ground state of Ce$^{3+}$ ion is a dipolar-octupolar (DO) Kramers doublet \cite{Sibille_Ce2Sn2O7, Gaudet_doQSI_dynamics, Gao_Ce2Zr2O7_2019, Sibille_doQSI_2020, Yao_Pi-flux_oQSI_2020, Bhardwaj_Ce2Zr2O7_Pi_flux, Smith_Ce2Zr2O7_Pi_flux, Desrochers_Ce2Zr2O7_2022, Hosoi_doQSI_2022}. The pseudospin-1/2 operator associated with this doublet features two pseudospin components that transform as dipole moments and one pseudospin component that transforms as an octupole moment \cite{Huang_doDoublets_2014, Li_do_doublets_2017}. 
Depending on the dominant pseudospin exchange constant, the subsequent Ising component dictates the ultimate phase of the underlying QSI, i.e.~dipolar (octupolar) QSI for a dipolar (octupolar) Ising pseudospin \cite{Patri_doQSI, Benton_do_pyrochlores_2020}.
Importantly, in the octupolar case, the emergent photon inherits the octupolar nature of the Ising component. 
The non-trivial symmetry nature of the octupolar moment provides a daunting task for inelastic neutron scattering, due to the lack of standard linear coupling with the neutron's dipolar moment.
Other promising QSI candidates, including $\rm{Pr}_2  \lk \rm{Zr}, \rm{Sn}, \rm{Hf} \rk_2 \rm{O}_7$, feature a non-Kramers doublet crystal-field ground state, which is only protected by the crystalline symmetries \cite{Kimura_Pr2Zr2O7_2013, Petit_Pr2Zr2O7_2016, Zhou_Pr2Sn2O7_2008, Princep_Pr2Sn2O7_2013, Sarte_Pr2Sn2O7_2017, Sibille_Pr2Hf2O7_2016, Anand_Pr2Hf2O7_2016, Sibille_Pr2Hf2O7_Photons_2018}. The Ising component transforms like a dipole moment whereas the transversal components transform like parts of a quadrupole.

As alluded to before, the transformation property of the Ising pseudospin component under lattice symmetry operations determines the form of the pseudospin-lattice coupling, which in turn determines the renormalization of the phonon spectrum and the speed of sound. It therefore suffices to consider two classes of QSI: ``dipolar-Ising'' and ``octupolar-Ising'' QSI. 
For example, octupolar (dipolar) QSI phases in $\rm{Ce}_2 \lk \rm{Sn}, \rm{Zr} \rk_2 \rm{O}_7$ correspond to ``octupolar-Ising" (``dipolar-Ising") QSI. The non-Kramers QSI in $\rm{Pr}_2  \lk \rm{Zr}, \rm{Sn}, \rm{Hf} \rk_2 \rm{O}_7$ corresponds to ``dipolar-Ising" QSI as the Ising component is dipolar.

Our study demonstrates that the renormalization of the phonon spectrum can be used to distinguish between dipolar-Ising and octupolar-Ising QSIs. 
In particular, due to the coupling of the emergent photon's gauge fields to the lattice degrees of freedom, the photon dynamics renormalize the low-energy phonon frequencies in particular ways depending on the examined high-symmetry momentum and magnetic field directions.
Though this renormalization is microscopically dependent on the coupling  between the photon and phonons, by comparing the ratio of renormalized phonon frequencies along different directions, we obtain coupling-independent predictions that may be verified in ultrasound studies.

The remainder of the paper is organized as follows: in Sec.~\ref{sec_mic_models} we describe the underlying microscopic models for the effective spin-1/2 and DO Kramers doublets as well as for the non-Kramers doublet. Then, in Sec.~\ref{sec_coupling} we introduce the precise form of the pseudospin-lattice couplings. In Sec.~\ref{sec_eff_action} we explain the effective low-energy theory in more detail. In particular, we calculate the corrections to the phonon action due to the emergent photons and derive the renormalization of the phonon spectrum.
Our results are presented in Sec.~\ref{sec_results}, where we also show how to extract the speed of the photons from the renormalized phonon spectrum.
We conclude with a brief discussion in Sec.~\ref{sec_discussion}.

\section{Microscopic Models}
\label{sec_mic_models}

\begin{figure}[t]

\subfloat[Pyrochlore lattice]{%
  \includegraphics[clip,width=0.48\columnwidth]{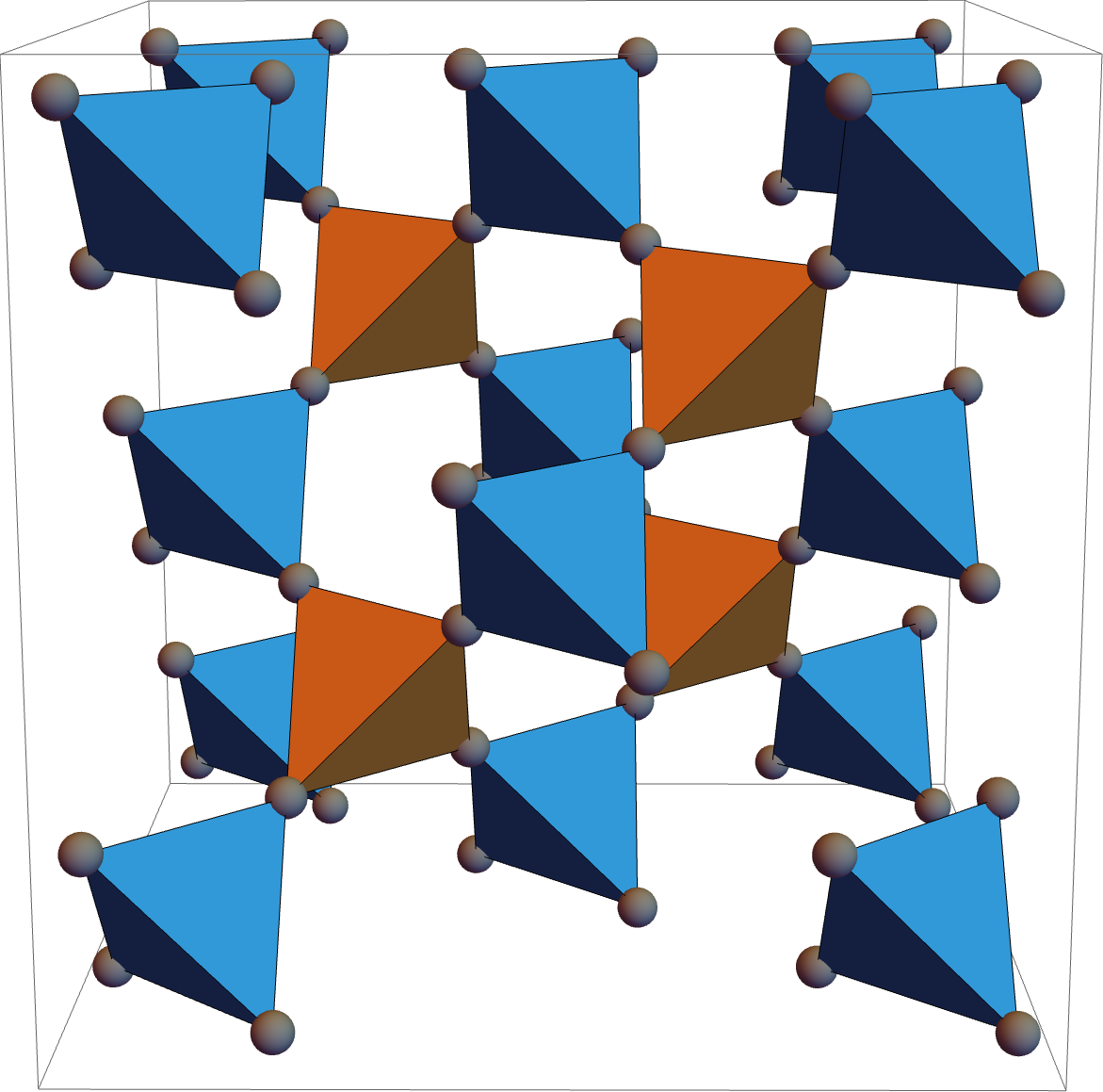}%
}
\hspace{0.5cm}
\subfloat[Local axes]{%
  \includegraphics[clip,width=0.45\columnwidth]{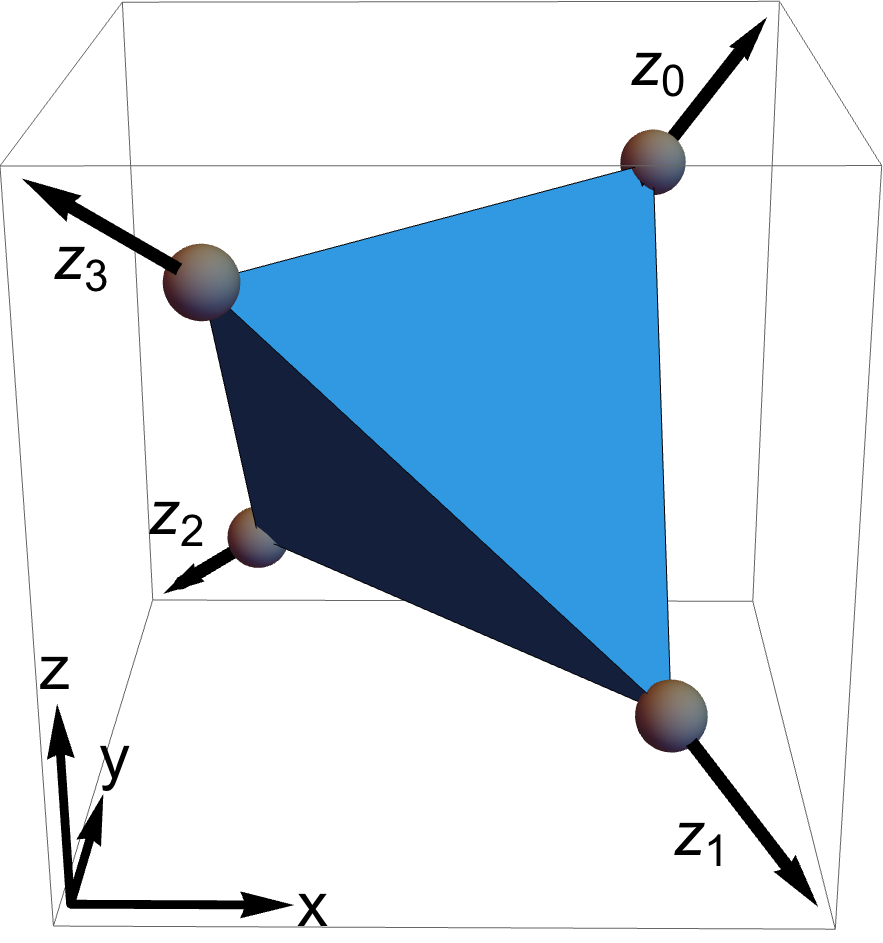}%
}

\caption{The pyrochlore lattice, shown in (a), is a FCC lattice with four sublattices. It forms a network of corner-sharing tetrahedra consisting of two types of tetrahedra (blue and orange), which differ only in their orientation. The rare-earth ions, located at the sites of the pyrochlore lattice, are depicted in grey. In (b) we show the local $z_{\alpha}$ axes for sublattices $\alpha \in \{0,1,2,3\}$ in relation to the global basis. The global basis vectors are $\hat{x} := (1,0,0)$, $\hat{y} := (0,1,0)$ and $\hat{z} := (0,0,1)$.}
\label{fig_pyrochlore}
\end{figure}

QSI arises in pyrochlore materials of the form $R_2 M_2 O_7$, where $R$ and $M$ refer to rare-earth and transition metal ions, respectively. The magnetic $R$ ions occupy the vertices of a network of corner-sharing tetrahedra, the pyrochlore lattice, which can be broken into four FCC sublattices as shown in Fig.~\ref{fig_pyrochlore}. Throughout the paper we often switch between a ``global'' frame coordinate system and ``local'' sublattice frames. The global frame refers to the standard Cartesian basis, see Fig.~\ref{fig_pyrochlore}. For each of the four sublattices, we define a different local basis as explained in Appendix \ref{app_glob_vs_loc}. The local $z$ axis is chosen such that it connects the centers of the two neighboring tetrahedra.

The magnetic property of the aforementioned compounds is dictated by the $f$ electrons of the $R$ ion. Spin-orbit coupling yields a degenerate set of states with total angular momentum $J$, whose degeneracy is partially lifted by the local $D_{3d}$ crystalline electric field (CEF) induced by the surrounding oxygen cage.
In the following, we assume that the CEF separates a ground state doublet sufficiently well from the higher lying energy levels, such that the low-energy analysis can be restricted to the subspace formed by the ground state doublet. This is indeed the case for the QSI candidates mentioned in the introduction \cite{Hodges_Yb2Ti2O7_2001, Sibille_Ce2Sn2O7, Gaudet_doQSI_dynamics, Kimura_Pr2Zr2O7_2013, Zhou_Pr2Sn2O7_2008, Princep_Pr2Sn2O7_2013, Sibille_Pr2Hf2O7_2016, Anand_Pr2Hf2O7_2016}. Each doublet can then be associated with a pseudospin-1/2 operator $\mathbf{S}$, which allows us to formulate the effective low-energy theory in terms of an interacting pseudospin Hamiltonian.

An odd number of $f$ electrons results in a Kramers doublet, whose degeneracy is protected by time reversal symmetry. Depending on the transformation properties under the $D_{3d}$ site symmetry (see Appendix \ref{app_d3d_sym}), we distinguish between two types of Kramers doublets. On the one hand, there is the effective spin-1/2 doublet, which transforms in the $\Gamma_4^+$ irreducible representation (irrep) of the $D_{3d}$ double group \cite{Huang_doDoublets_2014}. It can be found, for example, in $\rm{Yb}_2 \rm{Ti}_2 \rm{O}_7$ \cite{Rau_Frustrated_pyrochlore_2018}. All components of the corresponding pseudospin-1/2 operator transform like dipole components, see Appendix \ref{app_multipoles} for more details.

The most generic nearest-neighbor Hamiltonian for the effective spin-1/2 doublet reads
\begin{equation}
\begin{split}
\mathcal{H} =& \sum_{\langle i,j \rangle} \left[ J_{zz} S_z^i S_z^j - J_{\pm} \lk S_{+}^i S_{-}^j + S_{-}^i S_{+}^j \rk \right. \\
& \left. + J_{\pm \pm} \lk \beta_{ij} S_{+}^i S_{+}^j + \beta_{ij}^* S_{-}^i S_{-}^j \rk \right. \\
& \left. + J_{z \pm} \lk S_z^i \lk \zeta_{ij} S_{+}^j + \zeta_{ij}^* S_{-}^j \rk + \lk \zeta_{ij} S_{+}^i + \zeta_{ij}^* S_{-}^i \rk S_{z}^j \rk \right],
\end{split}
\label{generic_pseudo_ham}   
\end{equation}
where $S_z^i$ is the $z$ component of the pseudospin-1/2 operator written in the local basis of site $i$ (and similarly for the other components) \cite{Onoda_QSI_Ham_2011, Onoda_K+NK_Ham_2011, Savary_gMFT}. The sum runs over nearest neighbors and $J_{zz}$, $J_{\pm}$, $J_{\pm \pm}$ and $J_{z \pm}$ are coupling constants. Additionally, $\beta_{ij}$ and $\zeta_{ij} = -\beta_{ij}^*$ are unimodular complex numbers, see Appendix \ref{app_ham_const}. QSI arises in the frustrated regime, where $J_{zz} > 0$ and $J_{zz}~\gg~|J_{\pm}|, |J_{\pm \pm}|,|J_{z \pm}|$, for a certain range of $J_{\pm}$, $J_{\pm \pm}$ and $J_{z \pm}$, see e.g.~\cite{Savary_gMFT, Lee_Generic_QSI}.

On the other hand, as mentioned in the introduction, there also exists the possibility for a more exotic dipolar-octupolar (DO) Kramers doublet, which transforms as the direct sum of two one-dimensional irreps, $\Gamma_5^+ \oplus \Gamma_6^+$, of the $D_{3d}$ double group \cite{Huang_doDoublets_2014}. 
The QSI candidates $\rm{Ce}_2 \lk \rm{Sn}, \rm{Zr} \rk_2 \rm{O}_7$ support this type of doublet \cite{Patri_doQSI}.
Two of the pseudospin components, $S_x$ and $S_z$, transform like dipole components, whereas $S_y$ transforms like part of an octupole. More details can be found in Appendix \ref{app_multipoles}.
The symmetry transformation properties of the DO pseudospins allow us to rewrite the Hamiltonian in Eq.~(\ref{generic_pseudo_ham}) as the following XYZ model:
\begin{equation}
\begin{split}
\mathcal{H} =& \sum_{\langle i,j \rangle}  \mathcal{J}_{\mu} \tau_{\mu}^i \tau_{\mu}^j,
\end{split}
\label{ham_octupole}   
\end{equation}
where the sum over $\mu$ with $\mu \in \{x,y,z\}$ is implied and we introduce new pseudospin operators $\tau_{y} := S_y$, $\tau_x := \cos{(\theta)} S_x - \sin{(\theta)} S_z$ and $\tau_z := \sin{(\theta)} S_x + \cos{(\theta)} S_z$. The angle $\theta$ is determined by the exchange couplings $J_{ij}$ in Eq.~(\ref{generic_pseudo_ham}), see Appendix \ref{app_do_ham}.
We emphasize again that $ \tau_{\mu}^i$ is written in the local frame of site $i$. The new coupling constants $\mathcal{J}_{\mu}$ are combinations of the $J_{ij}$ constants from Eq.~(\ref{generic_pseudo_ham}) as shown in Appendix \ref{app_do_ham}.

If the number of $f$ electrons is even, a non-Kramers doublet can form, whose degeneracy is solely protected by the crystalline symmetries. It transforms as the $E_g$ irrep of the $D_{3d}$ point group and arises, for example, in $\rm{Pr}_2  \lk \rm{Zr}, \rm{Sn}, \rm{Hf} \rk_2 \rm{O}_7$ compounds \cite{Rau_Frustrated_pyrochlore_2018, Patri_qQSI}. The $S_x$ and $S_y$ pseudospin components associated with the non-Kramers doublet transform like components of a quadrupole, whereas $S_z$ transforms like a dipole (see Appendix \ref{app_multipoles}). 
We can use essentially the same Hamiltonian as for the effective spin-1/2 doublet in Eq.~(\ref{generic_pseudo_ham}). However, since $S_x$ and $S_y$ transform like time-reversal even quadrupole components, we must have $ J_{z \pm}=0$ to preserve time reversal symmetry.

\section{Pseudospin-lattice coupling}
\label{sec_coupling}

We now introduce pseudospin-lattice couplings to incorporate the elastic strain into the model. Ultimately, this allows us to derive the renormalization of the phonon spectrum due to the emergent photons.
Classically, for small deformations the elastic strain tensor $\epsilon$ is defined in the global frame as
\begin{equation}
    \epsilon_{jk} := \frac{1}{2}\lk \partial_j u_k + \partial_k u_j \rk.
\label{def_strain}
\end{equation}
$\mathbf{u}$ is the field describing the displacement of the atoms from equilibrium and $j,k \in \{x,y,z\}$ \cite{Landau_Elasticity}.
The free energy associated with the pseudospin-lattice coupling can be derived from representation theory arguments by imposing the relevant symmetries ($D_{3d}$ site symmetry and time reversal symmetry) and requiring that it transforms in the trivial representation. More details can be found in Appendix \ref{app_derive_coupling}.

Since we are interested in the photon contribution to the renormalization of the phonon spectrum, we only need to consider the coupling of the Ising pseudospin component to the elastic strain. To recapitulate, the reason for neglecting the transversal components is that they are associated with spinons, which are gapped excitations and hence do not contribute below the two-spinon-creation threshold \cite{Savary_gMFT}. 

Both, dipoles and octupoles, are time-reversal odd, meaning that the Ising pseudospin components of dipolar-Ising as well as octupolar-Ising QSI are also time-reversal odd. The elastic strain on the other hand is even under time reversal. Hence, linear coupling of the Ising component to the elastic strain requires the assistance of a time-reversal odd external magnetic field $\mathbf{h}$.

We first consider the case of octupolar-Ising QSI.
For octupolar-Ising QSI, $\tau_y = S_y$ is the Ising component and to lowest order, the free energy associated with the coupling takes the following form:
\begin{equation}
    \begin{split}
        \mathcal{F}_{oct} = 
        -  S_y^{\alpha} \left[ g_1 \lk 2 h_x^{\alpha} \epsilon_{xy}^{\alpha} +  h_y^{\alpha} \lk \epsilon_{xx}^{\alpha} - \epsilon_{yy}^{\alpha} \rk \rk \right. \\
        \left. + g_2 \lk h_y^{\alpha} \epsilon_{xz}^{\alpha} - h_x^{\alpha} \epsilon_{yz}^{\alpha} \rk \right],
    \end{split}
\label{coupling_oct}
\end{equation}
where $g_1$ and $g_2$ are phenomenological coupling constants \cite{Patri_doQSI}. All quantities indexed by $\alpha$ are written in the local frame of sublattice $\alpha$ and the sum over sublattices with $\alpha \in \{0,1,2,3\}$ is implied.

On the other hand, the Ising component for dipolar-Ising QSI is given by $S^z$, in which case the free energy corresponding to the coupling reads 
\begin{equation}
    \begin{split}
        \mathcal{F}_{dip} = -&\Tilde{g}_1 S_z^{\alpha} \left[ h_x^{\alpha}\lk \epsilon_{xx}^{\alpha} - \epsilon_{yy}^{\alpha} \rk   -2 h_y^{\alpha} \epsilon_{xy}^{\alpha}  \right]\\
        - &\Tilde{g}_2 S_z^{\alpha} \left[ h_x^{\alpha} \epsilon_{xz}^{\alpha} + h_y^{\alpha} \epsilon_{yz}^{\alpha}  \right]\\
        -&\Tilde{g}_3 S_z^{\alpha}  h_z^{\alpha} \left[ \epsilon_{xx}^{\alpha} + \epsilon_{yy}^{\alpha} \right] - \Tilde{g}_4 S_z^{\alpha}  h_z^{\alpha} \epsilon_{zz}^{\alpha},
    \end{split}
    \label{coupling_di}
\end{equation}
where $\Tilde{g}_1, \dots, \Tilde{g}_4$ are phenomenological coupling constants \cite{Patri_qQSI}. Again, quantities indexed by $\alpha$ are written in the local frame of sublattice $\alpha$ and the sum over sublattices $\alpha$ is implied.

We note that symmetry also allows a direct coupling of the Ising component to the external magnetic field, which is linear in the pseudospin. This is true for both octupolar-Ising and dipolar-Ising QSI. 
However, such a coupling merely leads to a constant shift in the effective action (see the next section) as long as the magnetic field is sufficiently small so that it does not cause a phase transition to a different field-induced state. This constant shift does not influence the phonon dynamics.
In principle, one could also include terms that are quadratic in the pseudospin and linear or quadratic in the strain, e.g.~$S_y^{\alpha} S_y^{\alpha} \epsilon_{zz}^{\alpha}$ for octupolar-Ising QSI. 
However, such terms do not affect the phonon spectrum up to $\mathcal{O}\lk h^2 \rk$. More specifically, if the correlator $\langle S_i^{\alpha} S_j^{\alpha} \rangle$ depends on the (external) magnetic field, then couplings of the form $ S_i^{\alpha} S_j^{\alpha}\epsilon_{mn}$ could give rise to magnetic-field-dependent corrections to the phonon action. Time reversal symmetry would force these to be at least of order $\mathcal{O}\lk h^4 \rk$ though. In this work we ignore such higher-order corrections as we only consider corrections up to $\mathcal{O}\lk h^2 \rk$.

\section{Effective low-energy theory}
\label{sec_eff_action}

The employment of a low-energy continuum description of the phonon dynamics encourages a similar continuum examination for the emergent photons. 
Indeed, as will be demonstrated in detail, such a low-energy theory incorporates the coupling between the lattice degrees of freedom and emergent photon, which ultimately renormalizes the phonon spectrum.

\subsection{Bare phonon spectrum}

Let us first consider the bare spectrum of acoustic phonons in the long-wavelength limit, which can be derived from the following imaginary time action:
\begin{equation}
    \mathcal{S}_{lat} =\int_{\tau, \textbf{r}} \left[  \frac{\rho}{2} \lk \partial_\tau \textbf{u}   \rk^2 +\mathcal{F}_s  \right].
\label{bare_phonon_action}
\end{equation}
$\rho$ is the mass density of the material, $\textbf{u}$ again denotes the displacement field
and $\mathcal{F}_s$ is the elastic energy of the underlying lattice \cite{Ye_Phonons_KitaevSL_2020}. As mentioned before, the pyrochlore lattice is a FCC lattice (with four sublattices). The corresponding point group, $O_h$, constraints the elastic energy $\mathcal{F}_s$ to be of the form
\begin{equation}
\begin{split}
    \mathcal{F}_s &= \frac{1}{2} c_{11} \lk \epsilon_{xx}^2 + \epsilon_{yy}^2 + \epsilon_{zz}^2 \rk \\
    &+ c_{12} \lk \epsilon_{xx}\epsilon_{yy}  + \epsilon_{xx}\epsilon_{zz}  + \epsilon_{yy}\epsilon_{zz}  \rk \\
    &+ 2  c_{44} \lk \epsilon_{xy}^2  + \epsilon_{xz}^2  + \epsilon_{yz}^2  \rk,
\end{split}
\end{equation}
where $c_{11}$, $c_{12}$ and $c_{44}$ are elastic constants in Voigt notation \cite{Luethi_Acoustics}.
We note that the elastic strain components are written in the global basis.

Fourier transforming the bare phonon action yields the bare inverse phonon propagator
\begin{widetext}
\renewcommand{\arraystretch}{1.5}
\begin{equation}
D^{-1} \lk \mathbf{q}, \omega_n \rk := 
    \begin{pmatrix}
    \rho \omega_n^2 + c_{11} q_x^2 + c_{44} \lk q_y^2 + q_z^2 \rk & \lk c_{12} + c_{44} \rk q_x q_y & \lk c_{12} + c_{44} \rk q_x q_z \\
    \lk c_{12} + c_{44} \rk q_x q_y & \rho \omega_n^2 + c_{11} q_y^2 + c_{44} \lk q_x^2 + q_z^2 \rk & \lk c_{12} + c_{44} \rk q_y q_z \\
    \lk c_{12} + c_{44} \rk q_x q_z & \lk c_{12} + c_{44} \rk q_y q_z & \rho \omega_n^2 + c_{11} q_z^2 + c_{44} \lk q_x^2 + q_y^2 \rk
    \end{pmatrix} ,
\label{secular_matrix}
\end{equation}
\end{widetext}
with momentum $\mathbf{q}$ and Matsubara frequency $\omega_n$. We can thus rewrite the action in Eq.~(\ref{bare_phonon_action}) as
\begin{equation}
    \mathcal{S}_{lat} = \sum_{\omega_n} \int_{\textbf{q}} \mathbf{u}^\top \lk -\mathbf{q}, -\omega_n \rk D^{-1} \lk \mathbf{q}, \omega_n \rk \mathbf{u} \lk \mathbf{q}, \omega_n \rk.
\end{equation}
The bare phonon spectrum can be obtained from the poles of the phonon Green's function by performing the analytic continuation $i\omega_n \longrightarrow \Omega + i 0^+$ and then solving $\text{det} \lk D^{-1} \rk  = 0$ for $\Omega$. In Sec.~\ref{sec_results} we show the renormalization of the phonon spectrum for certain high-symmetry momentum directions. The bare phonon spectrum for those high-symmetry momentum directions takes the form $\Omega^{(0)} = s ({\mathbf{q}}) |\mathbf{q}|$ (in the long-wavelength limit), where $s({\mathbf{q}})$ denotes the bare speed of sound, which depends on the specific momentum direction. Explicit expressions are given in Appendix \ref{app_speed_sound}.

\subsection{Magnetic-field-dependent corrections to the phonon action due to photons}

To obtain the renormalization of the phonon spectrum, we need to calculate the corrections to the bare phonon action arising from the interaction with the emergent photons.
We start by rewriting the photon action from Eq.~(\ref{S_QED}) in Matsubara frequency and momentum space:
\begin{widetext}
\begin{equation}
\mathcal{S}_{QED} = \frac{1}{2 K}\sum_{\omega_n} \int_{\textbf{q}} A_i(-\mathbf{q}, -\omega_n) \left[ \lk \omega_n^2 + v^2 q^2 \rk \delta_{ij} - v^2 q_i q_j \right]A_j(\mathbf{q}, \omega_n).
\label{S_QED_Fourier}
\end{equation}
\end{widetext}
The corresponding photon propagator then reads
\begin{equation}
    \begin{split}
        \langle A_i(-\mathbf{q}, -\omega_n) A_j(\mathbf{q}, \omega_n) \rangle &= \frac{K}{\omega_n^2 + v^2 q^2} \lk \delta_{ij}  + \frac{v^2 }{\omega_n^2}q_i q_j \rk\\
        &=: T_{ij}(\mathbf{q}, \omega_n).
    \end{split}
\label{photon_prop}
\end{equation}

Next, we have to find an expression for the pseudospin-lattice coupling in the low-energy subspace of the emergent photon. Let us recall that QSI can be regarded as an exponentially large superposition of classical spin ice (CSI) states \cite{Benton_Photons_QSI_2012}. These CSI states satisfy a local ``ice rule'' where the sum of the Ising components about each tetrahedron vanishes. Treating the local moments as quantum mechanical pseudospins, this becomes a zero-divergence constraint on the emergent electric field, which is defined on the links of the dual diamond lattice, i.e.~along the local $z$-axes. This implies that the local Ising component of the pseudospins should be mapped to the local $z$-component of the emergent electric field.
In the octupolar case, the Ising component is given by $S_y^{\alpha}$ and hence we have the mapping $S_y^{\alpha} \longrightarrow -\partial_\tau A_z^{\alpha} = E_z^{\alpha}$.
Using Eq.~(\ref{coupling_oct}) we then obtain the following imaginary time action associated with the coupling:
\begin{equation}
\begin{split}
    \mathcal{S}_{c}^{oct} = \int_{\tau, \textbf{r}} \sum_{\alpha} \lk \partial_{\tau}A_z^{\alpha}\rk &\left[ g_1 \lk 2  h_x^{\alpha} \epsilon_{xy}^{\alpha} + h_y^{\alpha} \lk \epsilon_{xx}^{\alpha} - \epsilon_{yy}^{\alpha} \rk \rk \right. \\
    &\left. + g_2 \lk h_y^{\alpha} \epsilon_{xz}^{\alpha} - h_x^{\alpha} \epsilon_{yz}^{\alpha} \rk \right].
\end{split}
\label{S_oct_coupling}
\end{equation}
As a reminder, all quantities indexed by $\alpha$ are written in the local frame of sublattice $\alpha$. We have to sum over all sublattices $\alpha$ and transform the local quantities to the global frame in order to find the renormalization of the phonon spectrum, which is expressed in the global frame.
The basis changes between local and global coordinates are described in Appendix \ref{app_glob_vs_loc}.
Going to momentum and Matsubara frequency space then yields
\begin{equation}
    \begin{split}
        \mathcal{S}_{c}^{oct} = \sum_{\omega_n} \int_{\textbf{q}} \omega_n   \mathbf{A}(-\mathbf{q}, -\omega_n) \cdot \mathbf{I}(\mathbf{q}, \omega_n),
    \end{split}
\label{S_oct_Fourier_glob}
\end{equation}
where $\mathbf{A}(\mathbf{q}, \omega_n)$ is the Fourier transform of the vector potential in the global frame and $\mathbf{I}(\mathbf{q}, \omega_n)$ encodes the (Fourier transformed) coupling of elastic strain and magnetic field in the global frame. The explicit form of $\mathbf{I}(\mathbf{q}, \omega_n)$ can be found in Appendix \ref{app_global_coupling}.

Since the total action,
\begin{equation}
    \mathcal{S}_{tot} =  \mathcal{S}_{lat} +  \mathcal{S}_{QED} +   \mathcal{S}_{c}^{oct},
\end{equation}
is quadratic in $\mathbf{A}$, the photons may be formally integrated out by completing the square to obtain an effective action only involving the phonons. This results in the following additional term
\begin{equation}
    \begin{split}
        \mathcal{S}_r^{oct} &= \sum_{\omega_n} \int_{\textbf{q}} \omega_n^2 \, \mathbf{I}^\top(-\mathbf{q},-\omega_n) T(\mathbf{q}, \omega_n)\mathbf{I}(\mathbf{q}, \omega_n),
    \end{split}
\label{renormalization_term_oct}
\end{equation}
which renormalizes the bare phonon action $\mathcal{S}_{lat}$. We use $T(\mathbf{q}, \omega_n)$ to denote the matrix associated with the photon propagator from Eq.~(\ref{photon_prop}).
Note that $\mathcal{S}_r$ is only quadratic in the displacement field $\mathbf{u}$ (since $\mathbf{I}(\mathbf{q}, \omega_n)$ is linear in $\mathbf{u}$ and $T(\mathbf{q}, \omega_n)$ does not depend on $\mathbf{u}$ at all). Therefore, we can rewrite the action in Eq.~(\ref{renormalization_term_oct}) as
\begin{equation}
    \begin{split}
        \mathcal{S}_r^{oct} = \sum_{\omega_n} \int_{\textbf{q}} \mathbf{u}^\top(-\mathbf{q},-\omega_n) \Pi(\mathbf{q}, \omega_n)\mathbf{u}(\mathbf{q}, \omega_n),
    \end{split}
\end{equation}
where $\Pi(\mathbf{q}, \omega_n)$ denotes the correction term due to the emergent photons.
We obtain the renormalized phonon spectrum from the poles of the (renormalized) phonon Green's function by performing the analytic continuation $i\omega_n \longrightarrow \Omega + i 0^+$ and then solving $\text{det} \lk D^{-1} + \Pi \rk  = 0$ for $\Omega$. Results for the renormalized phonon spectrum are presented in the next section where we also show how to extract the speed of the photons from it.

So far, we have discussed the phonon renormalization only for the octupolar-Ising QSI. However, the same strategy can be applied to the dipolar-Ising case. What changes is the coupling action which is now based on the pseudospin-lattice coupling from Eq.~(\ref{coupling_di}) and furthermore, $S_z^{\alpha} \longrightarrow -\partial_\tau A_z^{\alpha}$. The dipolar-Ising analogue of $\mathbf{I}(\mathbf{q},\omega_n)$, $\mathbf{\Tilde{I}}(\mathbf{q},\omega_n)$, is given in Appendix \ref{app_global_coupling}.

\section{Magnetic-field-dependent renormalization of the phonon spectrum}
\label{sec_results}

As shown in the previous section, the linear coupling of the emergent photons to the lattice degrees of freedom leads to a renormalized phonon action which is still quadratic in the displacement field. In effect, this leads to a renormalization of the speed of sound.
We present the renormalized phonon spectrum for the octupolar-Ising as well as dipolar-Ising QSI along various high-symmetry momentum and magnetic field directions in Table \ref{big_table}. 
The renormalized spectra are expanded in the limit $0 < \gamma_k h / s_i \ll 1$ ($0 < \gat_k h / s_i \ll 1$), $0 < v / s_i \ll 1$ and $ 0 < \gamma_k h \ll v$ ($ 0 < \gat_k h \ll v$) up to $\mathcal{O}\lk h^2 \rk$, where $h$ denotes the magnitude of the magnetic field, $s_i$ are bare speeds of sound in different momentum directions (listed in Appendix \ref{app_speed_sound}) and $v$ denotes again the speed of the emergent photon. $\gamma_k$ ($\gat_k$) are constants related to the pseudospin-lattice coupling constants $g_m$ ($\gt_m$), see Appendix \ref{app_gammas} for more details.
As is evident from Table \ref{big_table}, there is a strong orientation dependence for the renormalized speed of sound, where differing momentum and magnetic field configurations provide differing phonon renormalizations.
This strong directional dependence provides encouragement in discerning between the different classes of QSI based on measurements of the speed of sound.

\begin{table*}[ht]
\renewcommand{\arraystretch}{2}
\setlength{\tabcolsep}{8pt}
\centering
\caption{Summary of the phonon spectrum renormalization due to photons for octupolar-Ising QSI and dipolar-Ising QSI for different momentum and magnetic field directions. The bare speeds of sound, $s_i$, are given in Appendix \ref{app_speed_sound}. $\gamma_k$ and $\gat_k$ are constants composed of the respective pseudospin-lattice coupling constants, see Appendix \ref{app_gammas} for details. Due to the cubic symmetry, there are several equivalent combinations of momentum and magnetic field directions which yield the same solutions. For example, $(q,0,0)$ with $(h,0,0)$ is equivalent to $(0,0,q)$ with $(0,0,h)$. The table contains the renormalized spectra for several explicit, inequivalent combinations.}
\begin{tabular}{c|c|l|l}
\hline \hline
 $(q_x, q_y, q_z)$ & $(h_x, h_y, h_z)$ &  Octupolar-Ising: $\Omega_{\mathbf{q},\mathbf{h}}^{(j)}$ & Dipolar-Ising: $\Omega_{\mathbf{q},\mathbf{h}}^{(j)}$  \\ \hline
   
   \multirow{8}{*}{$(0,0,q)$} &\multirow{2}{*}{$(0,0,h)$} & $\Omega^{(1)} = s_1 |q| $ & $\Omega^{(1)} \simeq s_1 \left[ 1 - \frac{\gat_1}{s_1^2} h^2   \right] |q|  $\\
   & & $\Omega^{(2,3)}  \simeq s_2 \left[ 1 - \frac{\gamma_1}{s_2^2} h^2 \lk 1+ \frac{v^2}{s_2^2}  \rk  \right] |q|   $ & $\Omega^{(2,3)} \simeq s_2 \left[ 1 - \frac{\gat_2}{s_2^2} h^2 \lk 1+ \frac{v^2}{s_2^2}  \rk   \right] |q|  $  \\ \cline{2-4}
    
   &\multirow{3}{*}{$(0,h,0)$} & $\Omega^{(1)} \simeq s_1 \left[ 1 - \frac{\gamma_2}{s_1^2} h^2 \lk 1+ \frac{v^2}{s_1^2}  \rk   \right] |q|  $ &  $\Omega^{(1)} \simeq s_1 \left[ 1 - \frac{\gat_3}{s_1^2} h^2 \lk 1+ \frac{v^2}{s_1^2}  \rk   \right] |q|  $  \\
   & & $\Omega^{(2)} = s_2 |q| $ & $\Omega^{(2)} = s_2 |q| $\\
   & & $\Omega^{(3)} \simeq s_2 \left[ 1 - \frac{\gamma_1}{s_2^2} h^2   \right] |q|  $ & $\Omega^{(3)} \simeq s_2 \left[ 1 - \frac{\gat_2}{s_2^2} h^2  \right] |q|  $\\ \cline{2-4}
  
   &\multirow{3}{*}{$\dfrac{(h, h, 0)}{\sqrt{2}}$} & $\Omega^{(1)} \simeq s_1 \left[ 1 - \frac{\gamma_2}{s_1^2} h^2 \lk 1+ \frac{v^2}{s_1^2}  \rk  \right] |q|  $ & $\Omega^{(1)} \simeq s_1 \left[ 1 - \frac{\gat_3}{s_1^2} h^2 \lk 1+ \frac{v^2}{s_1^2}  \rk  \right] |q|  $ \\
   & & $\Omega^{(2)} = s_2 |q| $ & $\Omega^{(2)} = s_2 |q| $\\
   & & $\Omega^{(3)} \simeq s_2 \left[ 1 - \frac{\gamma_1}{s_2^2} h^2  \right] |q|  $ & $\Omega^{(3)} \simeq s_2 \left[ 1 - \frac{\gat_2}{s_2^2} h^2   \right] |q|  $ \\ \hline
  
   \multirow{9}{*}{$\dfrac{(q,q,0)}{\sqrt{2}}$} & \multirow{3}{*}{$(0,0,h)$} & $\Omega^{(1)} = s_3 |q| $ & $\Omega^{(1)} \simeq s_3 \left[ 1 - \frac{\gat_3}{s_3^2} h^2 \lk 1+ \frac{v^2}{s_3^2}  \rk  \right] |q|  $ \\ 
   &&  $\Omega^{(2)} \simeq s_2 \left[ 1 - \frac{\gamma_1}{s_2^2} h^2 \lk 1+ \frac{v^2}{s_2^2}  \rk  \right] |q|  $ & $\Omega^{(2)} \simeq s_2 \left[ 1 - \frac{\gat_2}{s_2^2} h^2   \right] |q|  $  \\
   && $\Omega^{(3)} \simeq s_4 \left[ 1 - \frac{\gamma_2}{s_4^2} h^2 \lk 1+ \frac{v^2}{s_4^2}  \rk  \right] |q| $ & $\Omega^{(3)} = s_4 |q| $ \\  \cline{2-4}
   
    & \multirow{3}{*}{$\dfrac{(h,h,0)}{\sqrt{2}}$} & $\Omega^{(1)} \simeq s_3 \left[ 1 - \frac{\gamma_3}{s_3^2} h^2 \lk 1+ \frac{v^2}{s_3^2}  \rk  \right] |q|  $ & $\Omega^{(1)} \simeq s_3 \left[ 1 - \frac{\gat_4}{s_3^2} h^2  \right] |q| $ \\ 
   &&  $\Omega^{(2)} \simeq s_2 |q|  $ &  $\Omega^{(2)} \simeq s_2 \left[ 1 - \frac{\gat_2}{s_2^2} h^2 \lk 1+ \frac{v^2}{s_2^2}  \rk  \right] |q|  $ \\
   && $\Omega^{(3)} \simeq s_4 \left[ 1 - \frac{1}{4}\frac{\gamma_2}{s_4^2} h^2  \right] |q| $ & $\Omega^{(3)} \simeq s_4 \left[ 1 - \frac{\gat_5}{s_4^2} h^2 \lk 1+ \frac{v^2}{s_4^2}  \rk  \right] |q|  $  \\  \cline{2-4}
   
   & \multirow{3}{*}{$\dfrac{(h,-h,0)}{\sqrt{2}}$} & $\Omega^{(1)} \simeq s_3 \left[ 1 - \frac{\gamma_4}{s_3^2} h^2  \right] |q|  $ &  $\Omega^{(1)} \simeq s_3 \left[ 1 - \frac{\gat_6}{s_3^2} h^2 \lk 1+ \frac{v^2}{s_3^2}  \rk  \right] |q|  $\\ 
   &&  $\Omega^{(2)} \simeq s_2 \left[ 1 - \frac{\gamma_1}{s_2^2} h^2 \lk 1+ \frac{v^2}{s_2^2}  \rk  \right] |q|  $ &  $\Omega^{(2)} \simeq s_2 |q|  $ \\
   && $\Omega^{(3)} \simeq s_4 \left[ 1 - \frac{1}{4}\frac{\gamma_2}{s_4^2} h^2 \lk 1+ \frac{v^2}{s_4^2}  \rk  \right] |q| $ & $\Omega^{(3)} \simeq s_4 \left[ 1 - \frac{\gat_5}{s_4^2} h^2   \right] |q|  $ \\  \hline
    
   \multirow{2}{*}{$\dfrac{(q,q,q)}{\sqrt{3}}$} &\multirow{2}{*}{$\dfrac{(h,h,h)}{\sqrt{3}}$} & $\Omega^{(1)} = s_5 |q| $ &  $\Omega^{(1)} \simeq s_5 \left[ 1 - \frac{\gat_7}{s_5^2} h^2  \right] |q|  $  \\
   && $\Omega^{(2,3)} \simeq s_6 \left[ 1 - \frac{\gamma_{5}}{s_6^2} h^2\lk 1 + \frac{v^2}{s_6^2} \rk  \right] |q|  $ & {$\Omega^{(2,3)} \simeq s_6 \left[ 1 - \frac{\gat_8}{s_6^2} h^2 \lk 1+ \frac{v^2}{s_6^2}  \rk  \right] |q|  $} \\ 
   \hline \hline
\end{tabular}
\label{big_table}
\end{table*}

\subsection{Extracting the speed of the photon from ratios of phonon spectra}

To extract the speed of the photons from the renormalized phonon spectrum, we calculate the ``renormalization ratio''
\begin{equation}
    \Delta^{(j)}_{[\mathbf{q},\mathbf{h}]} := -\frac{\Omega^{(j)}_{[\mathbf{q},\mathbf{h}]}-\Omega^{(0,j)}_{[\mathbf{q},\mathbf{h}]}}{\Omega^{(0,j)}_{[\mathbf{q},\mathbf{h}]}}
\end{equation}
for each renormalized solution $\Omega^{(j)}_{[\mathbf{q},\mathbf{h}]}$ separately. Here, $\mathbf{q}$ and $\mathbf{h}$ correspond to specific momentum and magnetic field directions, respectively.
For given $\mathbf{q}$ and $\mathbf{h}$, we get three, possibly degenerate, solutions which are labeled by $j$ (see Table \ref{big_table}). $\Omega^{(0,j)}_{[\mathbf{q},\mathbf{h}]}$ is the corresponding bare solution without the correction from the coupling of the emergent photons to the phonons. We use $\Tilde{\Delta}^{(j)}_{[\mathbf{q},\mathbf{h}]}$ to denote the ratio for the dipolar-Ising case in order to distinguish it from the octupolar-Ising ratio.

Though the renormalized spectra involve the coupling constants, one can consider different renormalization ratios, $\Delta^{(j)}_{[\mathbf{q},\mathbf{h}]}$ ($\Tilde{\Delta}^{(j)}_{[\mathbf{q},\mathbf{h}]}$), that involve the same $\gamma_k$ ($\gat_k$) coefficients, to obtain a pseudospin-lattice-coupling independent expression for the speed of the photon.

We now present two examples for octupolar-Ising QSI to illustrate the procedure. First, let us consider $\Delta^{(2,3)}_{[00q,00h]} = \frac{\gamma_1}{s_2^2} h^2 \lk 1+ \frac{v^2}{s_2^2}  \rk$ and $\Delta^{(3)}_{[00q,0h0]} = \frac{\gamma_1}{s_2^2} h^2$.
Taking the ratio of these and solving for $\frac{v}{s_2}$ yields
\begin{equation}
    \frac{v}{s_2} =  \sqrt{\dfrac{\Delta^{(2,3)}_{[00q,00h]}}{\Delta^{(3)}_{[00q,0h0]}} - 1},
\label{vs_oct}
\end{equation}
where $s_2$ denotes the bare transversal speed of sound in the $(0,0,q)$ direction, see Appendix \ref{app_speed_sound} for more details.
Note that the right-hand side is independent of any parameters.

Another renormalization ratio for octupolar-Ising QSI can be found by combining $\Delta^{(3)}_{[qq0,00h]} = \frac{\gamma_2}{s_4^2} h^2 \lk 1+ \frac{v^2}{s_4^2}  \rk$ and $\Delta^{(3)}_{[qq0,hh0]} = \frac{1}{4}\frac{\gamma_2}{s_4^2} h^2$. This leads to
\begin{equation}
    \frac{v}{s_4} =  \sqrt{\dfrac{1}{4}\dfrac{\Delta^{(3)}_{[qq0,00h]}}{\Delta^{(3)}_{[qq0,hh0]}} - 1},
\end{equation}
where $s_4$ denotes one of the bare transversal speeds of sound in the $(q,q,0)$ direction, see again Appendix \ref{app_speed_sound} for more details.

The above examples demonstrate that for weak but finite magnetic fields, i.e.~$0 < \gamma_k h \ll v$ ($0 < \gat_k h \ll v$), we can obtain the speed of the photons, $v$, without having precise knowledge of the pseudospin-lattice couplings $g_m$ ($\gt_m$).

\section{Discussion}
\label{sec_discussion}

In this work, we proposed ultrasound measurements as a tool for probing the emergent photons in QSI. We showed how these excitations renormalize the phonon spectrum (and associated speed of sound) of either octupolar-Ising or dipolar-Ising QSI. The latter includes the dipolar QSI arising in DO Kramers doublet compounds, the multipolar QSI phase of non-Kramers doublet materials, and the conventional (dipolar) QSI associated with effective spin-1/2 Kramers doublets.
Furthermore, we demonstrated how the speed of the emergent photon can be extracted from the renormalized spectrum without requiring knowledge of the precise effective coupling parameters. This protocol may help shed some light on the lowest energy excitations of QSI candidates such as $\rm{Ce}_2 \lk \rm{Sn}, \rm{Zr} \rk_2 \rm{O}_7$ \cite{Sibille_Ce2Sn2O7, Gaudet_doQSI_dynamics, Gao_Ce2Zr2O7_2019, Sibille_doQSI_2020} or $\rm{Pr}_2  \lk \rm{Zr}, \rm{Sn}, \rm{Hf} \rk_2 \rm{O}_7$ \cite{Kimura_Pr2Zr2O7_2013, Petit_Pr2Zr2O7_2016, Zhou_Pr2Sn2O7_2008, Princep_Pr2Sn2O7_2013, Sarte_Pr2Sn2O7_2017, Sibille_Pr2Hf2O7_2016, Anand_Pr2Hf2O7_2016, Sibille_Pr2Hf2O7_Photons_2018}.
Previous estimates from inelastic neutron scattering have suggested the speed of photons $v \approx \SI{3.6}{\meter/\second}$ for $\rm{Pr}_2  \rm{Hf}_2 \rm{O}_7$ \cite{Sibille_Pr2Hf2O7_Photons_2018}, while thermal conductivity measurements on $\rm{Pr}_2  \rm{Zr}_2 \rm{O}_7$ yield an estimate of $v \approx \SI{90}{\meter/\second}$ and a speed of sound $s \approx \SI{3700}{\meter/\second}$ \cite{Tokiwa_Thermal_conductivity_Photons}.
Using these estimates as a guide, it would be intriguing to discern the relative ratio of the speed of the emergent photons to the speed of the phonons of $v/s \sim 10^{-2} - 10^{-3}$ from our ultrasound measurement protocols.

Our results also suggest that measuring the phonon spectrum, or equivalently, the shift in the speed of sound, as a function of magnetic field should be sufficient in order to distinguish octupolar-Ising QSI from dipolar-Ising QSI. For example, the renormalization in the momentum direction $(0,0,q)$ with parallel magnetic field shows one unrenormalized solution for octupolar-Ising QSI, whereas all solutions in the dipolar-Ising case acquire a magnetic field dependence.
It would be interesting to see if our method, applied to DO doublet compounds such as $\rm{Ce}_2 \lk \rm{Sn}, \rm{Zr} \rk_2 \rm{O}_7$, could help determine the multipolar character of the QSI phase.

In terms of future work, it would be interesting to incorporate the other excitations of quantum spin ice (magnetic monopole as well as the spinon excitations) in this protocol. 
Such studies may prove to be fruitful in assisting investigations involving the detection of fractionalized excitations.

\begin{acknowledgments}
We thank an anonymous referee for pointing out an error in an earlier manuscript. This research was supported by the Natural Sciences and Engineering Research Council of Canada (NSERC) and the Centre for Quantum Materials at the University of Toronto.
\end{acknowledgments}

\appendix

\section{Global vs local bases}
\label{app_glob_vs_loc}

The pyrochlore lattice consists of four sublattices per unit cell. In Table \ref{local_basis} we list the local (orthonormal) basis vectors of each sublattice $\alpha \in \{0,1,2,3\}$ written in terms of the global basis vectors (i.e.~the standard Cartesian basis in 3d).
\begin{table}[ht]
\renewcommand{\arraystretch}{2}
\setlength{\tabcolsep}{2pt}
\centering
\caption{Local sublattice basis vectors in terms of the standard Cartesian basis \cite{Ross_QSI_Exp_2011}.}
\begin{tabular}{c c c c c}
\hline \hline
   $\alpha$ & 0 & 1 & 2 & 3  \\ \hline
   $\hat{x}^{\alpha}$ & $\frac{1}{\sqrt{6}} (-2,1,1)$ & $\frac{1}{\sqrt{6}} (-2,-1,-1)$ & $\frac{1}{\sqrt{6}} (2,1,-1)$ & $\frac{1}{\sqrt{6}} (2,-1,1)$\\
   $\hat{y}^{\alpha}$ & $\frac{1}{\sqrt{2}} (0,-1,1)$ & $\frac{1}{\sqrt{2}} (0,1,-1)$ & $\frac{1}{\sqrt{2}} (0,-1,-1)$ & $\frac{1}{\sqrt{2}} (0,1,1)$\\
   $\hat{z}^{\alpha}$ & $\frac{1}{\sqrt{3}} (1,1,1)$ & $\frac{1}{\sqrt{3}} (1,-1,-1)$ & $\frac{1}{\sqrt{3}} (-1,1,-1)$ & $\frac{1}{\sqrt{3}} (-1,-1,1)$\\
   \hline \hline
\end{tabular}
\label{local_basis}
\end{table}

Next, let us show how local physical quantities can be expressed in terms of the corresponding global quantities. For each sublattice $\alpha$ we need a different change of basis matrix $\mathcal{B}_{ \alpha}$, whose columns are the basis vectors from Table \ref{local_basis}, e.g.,
\begin{equation*}
    \begin{split}
        \mathcal{B}_{0} := 
        \begin{pmatrix}
           \frac{-2}{\sqrt{6}} & 0 & \frac{1}{\sqrt{3}}  \\
           \frac{1}{\sqrt{6}} & \frac{-1}{\sqrt{2}} & \frac{1}{\sqrt{3}} \\
           \frac{1}{\sqrt{6}} & \frac{1}{\sqrt{2}} & \frac{1}{\sqrt{3}}
        \end{pmatrix} .
    \end{split}
\end{equation*}

Vector quantities $\mathbf{V}^{\alpha}$, expressed in the local frame of sublattice $\alpha$, transform as follows: $\mathbf{V}^{\alpha} =  \mathcal{B}_{\alpha}^{-1} \mathbf{V} = \mathcal{B}_{\alpha}^{\top} \mathbf{V}$ where $ \mathbf{V}$ denotes the quantity in the global frame. Examples include the magnetic field $\mathbf{h}$, pseudospin-1/2 $\mathbf{S}$, momentum $\mathbf{q}$ and displacement field $\mathbf{u}$.

The elastic strain, a rank-2 tensor, has the following transformation behavior: $\epsilon^\alpha = \mathcal{B}_{\alpha}^{-1} \epsilon \mathcal{B}_{\alpha} = \mathcal{B}_{\alpha}^{\top} \epsilon \mathcal{B}_{\alpha}$.

\section{Local symmetry transformations}
\label{app_d3d_sym}

The $D_{3d}$ point group can be obtained from the following two generators:
\begin{equation*}
\mathcal{S}^-_6 := 
    \begin{pmatrix}
     \frac{1}{2} & \frac{\sqrt{3}}{2} & 0  \\
     -\frac{\sqrt{3}}{2} & \frac{1}{2} & 0 \\
     0 & 0 & -1
    \end{pmatrix} \hspace{0.5cm}
    \mathcal{C}^{'}_{21} := 
    \begin{pmatrix}
     -1 & 0 & 0  \\
     0 & 1 & 0 \\
     0 & 0 & -1
    \end{pmatrix},
\label{sym_generators}
\end{equation*}
which refer to the local bases. $\mathcal{S}^-_6$ is an improper rotation about the (local) $\hat{z}$ and $\mathcal{C}^{'}_{21}$ is a $\pi$ rotation about the (local) $\hat{y}$ axis.

The local Ising pseudospin component in the octupolar case, $\tau_y^{\alpha}$, transforms as follows \cite{Patri_doQSI}:
\begin{equation}
    \begin{split}
        \mathcal{S}^-_6 &: \tau_y^{\alpha} \longrightarrow \tau_y^{\alpha} \\
        \mathcal{C}^{'}_{21} &: \tau_y^{\alpha} \longrightarrow \tau_y^{\alpha} .
    \end{split}
\end{equation}

Similarly, in the dipolar-Ising case, we obtain the following transformation behavior for $S_z^{\alpha}$ \cite{Patri_qQSI}:
\begin{equation}
    \begin{split}
        \mathcal{S}^-_6 &: S_z^{\alpha} \longrightarrow S_z^{\alpha} \\
        \mathcal{C}^{'}_{21} &: S_z^{\alpha} \longrightarrow -S_z^{\alpha}.
    \end{split}
\end{equation}

Regarding symmetry transformations of the magnetic field, we need to recall that the magnetic field transforms like a pseudovector, meaning that it picks up an additional minus sign under improper rotations like $\mathcal{S}^-_6$. 

Finally, the strain tensor transforms as $\epsilon \longrightarrow M \epsilon M^{-1}$ where $M$ denotes the symmetry element.

\section{Pseudospin-1/2 in terms of multipole operators}
\label{app_multipoles}

As argued in the main text, a given ground state doublet can be represented by a pseudospin-1/2 operator $\mathbf{S}$. Depending on the symmetry transformation properties of the doublet, the pseudospin-1/2 components can be expressed in terms of different multipole operators. Using the Wigner-Eckart theorem one can relate multipole operators to certain combinations of the total angular momentum components $J_x$, $J_y$ and $J_z$ \cite{Kuramoto_Multipoles, Kusunose_Multipoles}. 

For the effective spin-1/2 Kramers doublet one finds
\begin{equation}
    \begin{split}
        S_x &\propto P J_x P \\
        S_y &\propto P J_y P \\
        S_z &\propto P J_z P,
    \end{split}
\end{equation}
where $P$ denotes the projector onto the ground state doublet \cite{Onoda_QSI_Ham_2011}.

In the case of the DO Kramers doublet we have
\begin{equation}
    \begin{split}
        S_x &= P \lk c_0 \lk J_x^3 - \overline{J_x J_y J_y}\rk + c_1 J_z \rk P \\
        S_y &= P c_2 \lk J_y^3 - \overline{J_y J_x J_x}\rk P \\
        S_z &= P c_3 J_z P.
    \end{split}
\end{equation}
$P$ now projects onto the DO Kramers ground state doublet and the overline indicates a symmetrized product \cite{Patri_doQSI}.
$c_0, \dots, c_3$ are phenomenological constants depending on the CEF parameters. They are needed to ensure that the pseudospin-1/2 components at a specific site satisfy $\left[ S_j, S_k \right] = i \epsilon_{jkl} S_l$ where $j,k,l \in \{x,y,z\}$ (i.e.~they are a basis of the $su(2)$ algebra). For $\rm{Ce}_2 \lk \rm{Sn}, \rm{Zr} \rk_2 \rm{O}_7$ it can be shown that $c_1 = 0$ \cite{Patri_doQSI}. Despite $S_x$ featuring an octupolar operator, $J_x^3 - \overline{J_x J_y J_y}$, we refer to it as being dipolar since it transforms identically to the dipole component $J_z$.

The pseudospin-1/2 components for the non-Kramers doublet in $\rm{Pr}_2 \rm{Zr}_2 \rm{O}_7$ can be represented as \cite{Patri_qQSI}
\begin{equation}
    \begin{split}
        S_x &\propto P \overline{J_x J_z} P\\
        S_y &\propto P \overline{J_y J_z} P\\
        S_z &\propto P J_z P,
    \end{split}
\end{equation}
where $P$ now is the projector onto the non-Kramers doublet. We note that $S_x$ and $S_y$ transform like quadrupole components.

\section{Constants of the pseudospin-1/2 Hamiltonian}
\label{app_ham_const}

For the generic pseudospin-1/2 Hamiltonian in Eq.~(\ref{generic_pseudo_ham}) we use the following $\beta$ matrix:
\begin{equation*}
\beta :=
    \begin{pmatrix}
     0 & 1 & w & w^2 \\
     1 & 0 & w^2 & w \\
     w & w^2 & 0 & 1 \\
     w^2 & w & 1 & 0 
    \end{pmatrix} ,
\label{beta_matrix}
\end{equation*}
where $w:=e^{2 \pi i/ 3}$ \cite{Savary_gMFT}.

\section{Hamiltonian for the DO doublet}
\label{app_do_ham}

We start with the generic pseudospin-1/2 Hamiltonian in Eq.~(\ref{generic_pseudo_ham}).
It can be shown that for the DO doublet, $w = 1$ in Eq.~(\ref{beta_matrix}). Hence, the effective Hamiltonian can be rewritten as
\begin{equation}
\begin{split}
\mathcal{H} =& \sum_{\langle i,j \rangle} \left[ J_{xx} S_x^i S_x^j + J_{yy} S_y^i S_y^j \right. \\
& \left.+ J_{zz} S_z^i S_z^j + J_{xz}(S_x^i S_z^j + S_z^i S_x^j) \right],
\end{split}
\end{equation}
where $J_{xx} := 2(J_{\pm \pm} - J_{\pm})$, $J_{yy} := -2(J_{\pm \pm} + J_{\pm})$ and $J_{xz} := 2J_{z \pm})$ \cite{Rau_Frustrated_pyrochlore_2018, Huang_doDoublets_2014}. Since $S_x$ and $S_z$ both transform in the $\Gamma_2^+$ irrep of the $D_{3d}$ double group, their coupling in the above Hamiltonian can be eliminated by a rotation in pseudospin space
\begin{align*}
        \tau_x &:= \cos{(\theta)} S_x - \sin{(\theta)} S_z \\
        \tau_{y} &:= S_y \\
        \tau_z &:= \sin{(\theta)} S_x + \cos{(\theta)} S_z ,    
\end{align*}
where $\theta := \frac{1}{2} \arctan{\lk \frac{2 J_{xz}}{J_{zz} - J_{xx}} \rk}$.
This leads to the desired Hamiltonian in Eq.~(\ref{ham_octupole}) with
\begin{align*}
        \mathcal{J}_x &:= \frac{1}{2} \lk J_{xx} + J_{zz} - \sqrt{(J_{zz} - J_{xx})^2 + 4 J_{xz}^2} \rk \\
        \mathcal{J}_y &:= J_{yy} \\
        \mathcal{J}_z &:= \frac{1}{2} \lk J_{xx} + J_{zz} + \sqrt{(J_{zz} - J_{xx})^2 + 4 J_{xz}^2} \rk .    
\end{align*}

\section{Derivation of the pseudospin-lattice coupling}
\label{app_derive_coupling}

The pseudospin-lattice coupling can be derived from representation theory arguments as follows: since the free energy should be time-reversal even, we only need to consider combinations of pseudospin-1/2, magnetic field and elastic strain components that are even under time reversal. Furthermore, the free energy, a scalar, has to transform in the trivial representation of the $D_{3d}$ point group. We can then construct a projection operator $\mathcal{P}$,
\begin{equation}
    \mathcal{P} := \frac{1}{d}\sum_M \chi(M) O(M),
\end{equation}
which projects any given combination of pseudospin-1/2, magnetic field and elastic strain components onto the subspace of the trivial representation \cite{Tinkham}. Here, $d=12$ is the order of the $D_{3d}$ point group and the sum runs over all group elements $M$. Generators of the $D_{3d}$ point group are given in in Appendix \ref{app_d3d_sym}.
The characters of the trivial representation satisfy $\chi(M) = 1$ for all $M$ and $O(M)$ acts on a given function (i.e.~combination of pseudospin-1/2, magnetic field and elastic strain components) by performing the symmetry transformation specified by the group element $M$.
For example, in the octupolar-Ising case we obtain $\mathcal{P}(S_y^{\alpha} h_y^{\alpha} \epsilon_{xx}^{\alpha}) \propto 2 h_x^{\alpha} \epsilon_{xy}^{\alpha} +  h_y^{\alpha} \lk \epsilon_{xx}^{\alpha} - \epsilon_{yy}^{\alpha} \rk$ but $\mathcal{P}(S_y^{\alpha} h_z^{\alpha} \epsilon_{xx}^{\alpha}) = 0$.

\section{Bare speeds of sound}
\label{app_speed_sound}

The bare speeds of sound in the $(q,0,0)$, $(0,q,0)$ and $(0,0,q)$ momentum directions are given by
\begin{align*}
        s_1 &:= \sqrt{\frac{c_{11}}{\rho}}\\
        s_2 &:= \sqrt{\frac{c_{44}}{\rho}}, 
\end{align*}
where $s_1$ is associated with the longitudinal and $s_2$ with the transversal modes \cite{Luethi_Acoustics}.

In addition to $s_2$, the $(0,q,q)$, $(q,0,q)$ and $(q,q,0)$ directions come with
\begin{align*}
        s_3 &:= \sqrt{\frac{c_{11} + 2 c_{44} + c_{12}}{2\rho}}\\
        s_4 &:= \sqrt{\frac{c_{11} - c_{12}}{2\rho}}. 
\end{align*}
Here, $s_3$ corresponds to the longitudinal mode, whereas $s_2$ and $s_4$ correspond to transversal modes.

Finally, for $(q,q,q)$ we have
\begin{align*}
        s_5 &:= \sqrt{\frac{c_{11} +  4 c_{44} + 2 c_{12}}{3\rho}}\\
        s_6 &:= \sqrt{\frac{c_{11} + c_{44} - c_{12}}{3\rho}},     
\end{align*}
with $s_5$ being the longitudinal and $s_6$ the transversal speed of sound.

\section{Fourier transform of the pseudospin-lattice coupling in the global frame}
\label{app_global_coupling}

In order to calculate the phonon self-energy due to photons, we Fourier transformed the pseudospin-lattice couplings (Eq.~(\ref{coupling_oct}) and Eq.~(\ref{coupling_di})) and additionally expressed them in the global basis. For notational convenience, we then introduced vector-like quantities $\mathbf{I}(\mathbf{q},\omega_n)$ and $\mathbf{\Tilde{I}}(\mathbf{q},\omega_n)$, which encode only the coupling of the magnetic field and the lattice degrees of freedom.
The explicit form of $\mathbf{I}(\mathbf{q},\omega_n)$ for octupolar-Ising QSI is given by
\begin{align*}
        I_x(\mathbf{q}, \omega_n) &:= \eta_1 (h_y q_y - h_z q _z) u_x(\mathbf{q}, \omega_n) \\
        & + \lk \eta_1 h_y q_x + \eta_2 h_x q_y \rk u_y(\mathbf{q}, \omega_n) \\
        & - \lk \eta_1 h_z q_x + \eta_2 h_x q_z \rk u_z(\mathbf{q}, \omega_n) \\
        I_y(\mathbf{q}, \omega_n) &:= -\lk \eta_2 h_y q_x + \eta_1 h_x q_y \rk u_x(\mathbf{q}, \omega_n) \\
        & + \eta_1 (h_z q_z - h_x q _x) u_y(\mathbf{q}, \omega_n) \\
        & + \lk \eta_1 h_z q_y + \eta_2 h_y q_z \rk u_z(\mathbf{q}, \omega_n) \\
        I_z(\mathbf{q}, \omega_n) &:= \lk \eta_2 h_z q_x + \eta_1 h_x q_z \rk u_x(\mathbf{q}, \omega_n) \\
        & -\lk \eta_2 h_z q_y + \eta_1 h_y q_z \rk u_y(\mathbf{q}, \omega_n) \\
        & + \eta_1 (h_x q_x - h_y q _y) u_z(\mathbf{q}, \omega_n) ,   
\end{align*}
where
\begin{align*}
        \eta_1 & := \frac{2}{3 \sqrt{3}} \lk 2 \sqrt{2} g_1 +  g_2 \rk\\
        \eta_2 & := \frac{4}{3 \sqrt{3}} \lk \sqrt{2} g_1 - g_2 \rk.
\end{align*}

For dipolar-Ising QSI, on the other hand, we define
\begin{align*}
        \Tilde{I}_x(\mathbf{q}, \omega_n) &:= \lk -\et_1 h_x q_x + \et_2 h_y q_y + \et_2 h_z q_z  \rk u_x(\mathbf{q}, \omega_n) \\
        & + \lk \et_2 h_y q_x + \et_3 h_x q_y \rk u_y(\mathbf{q}, \omega_n) \\
        & + \lk \et_2 h_z q_x + \et_3 h_x q_z \rk u_z(\mathbf{q}, \omega_n) \\
        \Tilde{I}_y(\mathbf{q}, \omega_n) &:= \lk \et_3 h_y q_x + \et_2  h_x q_y  \rk u_x(\mathbf{q}, \omega_n) \\
        & + \lk \et_2 h_x q_x - \et_1 h_y q_y + \et_2 h_z q_z \rk u_y(\mathbf{q}, \omega_n) \\
        & + \lk \et_3 h_z q_y + \et_2 h_y q_z \rk u_z(\mathbf{q}, \omega_n) \\
        \Tilde{I}_z(\mathbf{q}, \omega_n) &:= \lk \et_3 h_z q_x + \et_2  h_x q_z \rk u_x(\mathbf{q}, \omega_n) \\
        & + \lk \et_3 h_z q_y + \et_2 h_y q_z \rk u_y(\mathbf{q}, \omega_n) \\
        & + \lk \et_2 h_x q_x + \et_2 h_y q_y - \et_1 h_z q_z \rk u_z(\mathbf{q}, \omega_n) ,     
\end{align*}
where
\begin{align*}
       \et_1 & := \frac{4}{9} \lk 2 \sqrt{2} \gt_1 - 2\gt_2 - 2\gt_3 - \gt_4 \rk\\
       \et_2 & := \frac{2}{9} \lk 2 \sqrt{2} \gt_1 + \gt_2 - 2\gt_3 + 2\gt_4 \rk\\
       \et_3 & := \frac{4}{9} \lk \sqrt{2} \gt_1 - \gt_2 + 2\gt_3 + \gt_4 \rk.
\end{align*}

\section{Constants in the phonon spectrum}
\label{app_gammas}

Here, we list all the coupling constants $\gamma_k$ and $\gat_k$ appearing in Table \ref{big_table} in terms of the pseudospin-lattice couplings $g_m$ and $\gt_m$.

For octupolar-Ising QSI we have
\begin{align*}
        \gamma_1 & := \frac{2}{27}\frac{K}{\rho} \lk 2 \sqrt{2} g_1 + g_2 \rk^2\\
        \gamma_2 & := \frac{8}{27}\frac{K}{ \rho} \lk \sqrt{2} g_1 - g_2 \rk^2\\
        \gamma_3 & := \frac{4}{3}\frac{K}{ \rho} g_1^2\\
        \gamma_4 & := \frac{4}{27}\frac{K}{ \rho} \lk g_1 + \sqrt{2}  g_2 \rk^2\\
        \gamma_5 & := \frac{2}{81}\frac{K}{ \rho} \lk 4 \sqrt{2} g_1 - g_2 \rk^2,
\end{align*}

where $K$ is the phenomenological constant appearing in the effective QED action in Eq.~(\ref{S_QED}) and $\rho$ denotes the mass density of the material.

In the dipolar-Ising case we define
\begin{align*}
        \gat_1 & := \frac{8}{81}\frac{K}{ \rho} \lk 2 \sqrt{2} \gt_1 - 2\gt_2 - 2\gt_3 - \gt_4 \rk^2\\
        \gat_2 & := \frac{2}{81}\frac{K}{ \rho} \lk 2 \sqrt{2} \gt_1 + \gt_2 - 2\gt_3 + 2\gt_4 \rk^2\\
        \gat_3 & := \frac{8}{81}\frac{K}{\rho} \lk \sqrt{2} \gt_1 - \gt_2 + 2\gt_3 + \gt_4 \rk^2\\
        \gat_4 & := \frac{4}{81}\frac{K}{ \rho} \lk \gt_1 + \sqrt{2} \gt_2 + \sqrt{2}\gt_3 + 2 \sqrt{2}\gt_4 \rk^2\\
        \gat_5 & := \frac{2}{9}\frac{K}{\rho} \lk \sqrt{2} \gt_1 - \gt_2 \rk^2\\
        \gat_6 & := \frac{4}{9}\frac{K}{\rho} \lk \gt_1 - \sqrt{2}\gt_3 \rk^2\\
        \gat_7 & := \frac{8}{729}\frac{K}{ \rho} \lk 4\sqrt{2} \gt_1 + 2\gt_2 + 2\gt_3 + 7\gt_4 \rk^2\\
        \gat_8 & := \frac{2}{729}\frac{K}{ \rho} \lk 4\sqrt{2} \gt_1 - 7\gt_2 + 2\gt_3 - 2\gt_4 \rk^2.
\end{align*}

\bibliography{References}

\end{document}